\newcommand{\PRL}[3]{Phys.\ Rev.\ Lett.\ {\bf #1},\ #2 (#3)}
\newcommand{\RMP}[3]{Rev.\ Mod.\ Phys.\ {\bf #1},\ #2 (#3)}
\newcommand{\IEEE}[3]{IEEE\ J.\ Sel.\ Top.\ Quantum \ Electron.\ {\bf #1},\ #2 (#3)}
\newcommand{\OL}[3]{Opt.\ Lett.\ {\bf #1},\ #2 (#3)}
\newcommand{\NAT}[3]{Nature\ {\bf #1},\ #2 (#3)}
\newcommand{\SC}[3]{Science\ {\bf #1},\ #2 (#3)}

\newcommand{\PRA}[3]{Phys.\ Rev.\ A\ {\bf #1},\ #2 (#3)}

\newcommand{\PRE}[3]{Phys.\ Rev.\ E\ {\bf #1},\ #2 (#3)}

\newcommand{\JPB}[3]{J.\ Phys.\ B:\ At.\ Mol.\ Opt.\ Phys.\ {\bf #1},\ #2 (#3)}
\newcommand{\JPA}[3]{J.\ Phys.\ A:\ Math.\ Gen.\ {\bf #1},\ #2 (#3)}

\newcommand\sech{\mathrm{sech}}

\newcommand{\SCR}{Schr\"odinger~}
\newcommand{\GP}{Gross-Pitaevskii~}
\newcommand{\diracslash}[1]{#1\llap{/\kern2pt}}

\newcommand{\be}{\begin{equation}}
\newcommand{\ee}{\end{equation}}
\newcommand{\bea}{\begin{eqnarray}}
\newcommand{\eea}{\end{eqnarray}}
\newcommand{\ba}[1]{\begin{array}{#1}}
\newcommand{\ea}{\end{array}}

\newcommand{\eqrf}[1]{Eq.\ (\ref{#1})}

\documentclass[twocolumn,superscriptaddress,secnumarabic,amssymb,nobibnotes,nofootinbib,aps,prl,showpacs]{revtex4}
\usepackage{graphicx}
\begin{document}
\setlength{\topmargin}{-0.05in}

\title{A new class of exact solitary wave solutions of one dimensional \GP equation}

\author{Rajneesh Atre}\thanks{\texttt{atre@prl.res.in}}
 \affiliation{ Physical
Research Laboratory, Navrangpura, Ahmedabad 380 009, India}
 \affiliation{
School of Physics, University of Hyderabad, Hyderabad-500 046,
India}

\author{Prasanta K. Panigrahi}\thanks{\texttt{prasanta@prl.res.in}}
\affiliation{ Physical Research Laboratory, Navrangpura, Ahmedabad
380 009, India}

\author{G.S. Agarwal}\thanks{\texttt{girish.agarwal@okstate.edu}}\affiliation{Department of Physics,
Oklahoma State University, Stillwater, OK 74078,USA}\affiliation
{Physical Research Laboratory, Navrangpura, Ahmedabad 380 009,India}

\date{\today}

\begin{abstract}
We present a large family of {\it{exact}} solitary wave solutions of
the one dimensional \GP equation, with time-varying scattering
length and gain/loss, in both expulsive and regular parabolic
confinement regimes. The consistency condition governing the soliton
profiles is shown to map on to a {\it{linear}} \SCR eigenvalue
problem, thereby enabling one to find analytically the effect of a
wide variety of temporal variations in the control parameters, which
are experimentally realizable. Corresponding to each solvable
quantum mechanical system, one can identify a soliton configuration.
These include soliton trains in close analogy to experimental
observations of Strecker {\it{et al.,}} [\NAT {417}{150}{2002}],
spatio-temporal dynamics, solitons undergoing rapid amplification,
collapse and revival of condensates and analytical expression of
two-soliton bound states, to name a few.

\end{abstract}


\maketitle Coherent atom optics is the subject of much current
interest due to its relevance to both fundamental aspects of
physics, as well as to technology
\cite{Lenz,Cornell,Kett,Busch,Gupta,Bloch}. For that purpose, lower
dimensional condensates {\em{e.g.,}} cigar-shaped Bose-Einstein
condensates (BECs) have been the subject of active study in the last
few years \cite{Garcia,Stringari,Carr1,Liang,Konotop}. Observations
of dark and bright solitons \cite{Burger,Khaykovich,Strecker},
particularly the latter one, since the same is a condensate itself,
have generated considerable interest in this area. This has spurred
intense investigations about the behavior of condensates in the
presence of time-varying control parameters. These include
nonlinearity, achievable through Feshbach resonance
\cite{Vogels,Kett1,Wieman}, gain/loss and the oscillator frequency
\cite{Bigelow}. The fact that for a condensate in oscillator
potential, exact solutions of the \GP (GP) equation are not
available, makes it extremely difficult to examine the effects of
time variation in the aforementioned parameters. In the context of
pulse propagation in non-linear optical fibers, a number of authors
have recently investigated the effects of variable non-linearity,
dispersion and gain or loss. Moores analyzed this problem for
constant dispersion and nonlinearity and a distributed gain
\cite{Moores}, whereas Kruglov {\em{et al.,}} considered the same
problem with all the parameters in a variable form \cite{Kruglov}.
Serkin {\it{et al.,}} have derived the nonlinear \SCR equation
(NLSE) with distributed coefficients as compatibility condition of
two first order equations, demonstrating the applicability of
inverse scattering transform method to this type of problems
\cite{HaseIEEE}. They write down the general equation relating the
distributed coefficients with the solution parameters and obtain a
number of exact solutions. Recently, exact solutions of a driven
NLSE with distributed dispersion, nonlinearity and gain, which
exhibits pulse compression in a twin-core optical fiber, has also
been obtained \cite{PKP1}.

In this Letter, we present a large family of exact solutions of the
quasi one-dimensional GP equation, which is the familiar NLSE, with
time varying scattering length, gain/loss, in the presence of an
oscillator potential, which can be both expulsive or regular. It is
shown that, the consistency condition governing the soliton profiles
identically maps on to the {\it{linear}} \SCR eigenvalue problem,
thereby allowing one to solve analytically the GP equation for a
wide variety of temporal variations in the control parameters.
Corresponding to each solvable quantum mechanical eigenvalue
problem, one can identify a soliton-like profile. These can be dark
or bright and the oscillator frequencies can have a variety of
temporal profiles, including linear, quadratic, exponential, smooth
step-like potential and kicked oscillator scenario. Our solutions
exhibit soliton trains, spatio-temporal dynamics of solitons, the
formation of two-soliton bound states, to name a few. Interestingly,
it was found that some of these solitons can periodically exchange
atoms with a background. Further, our analytical results are closely
related to experiments \cite{Khaykovich,Strecker}. Amplification of
atomic condensate and condensate compression are observed in the
parameter domain, which are amenable for experimental verification.

Analogous to the experimental observations of Strecker {\it{et
al.,}} we obtain bright soliton trains in the presence of harmonic
confinement \cite{Strecker}. Some of the soliton solutions in the
presence of regular harmonic confinement, with appropriately
tailored gain/loss, exhibit collapse and revival phenomena, with an
increase in amplitude. We treat the attractive regime rather
exhaustively since the  bright solitons are themselves condensates.
The analytic expression for the dark soliton in the repulsive sector
has been presented, detailed analysis of the same can be carried out
in a straightforward manner.

We start with a zero temperature BEC of atoms, confined in
cylindrical harmonic trap
$V_{0}(x,y)=m\omega_{\perp}^{2}(x^2+y^2)/2$ and a time-dependent
harmonic confinement, which can be both attractive and expulsive,
along the $z$-direction $V_{1}(z,t)=m\omega_{0}^{2}(t)z^2/2$:

\be \label{GP} i\hbar \frac{\partial \Psi({\bf{r}},t)}{\partial
t}=\left\{-\frac{\hbar^2}{2m}\nabla^{2}+U|\Psi({\bf{r}},t)|^{2}+
V+i\frac{\eta(t)}{2}\right\}\Psi({\bf{r}},t),\ee where
$U=4\pi\hbar^2a_{s}(t)/m$ and $V=V_{0}+V_{1}$. It is worth noting
that, condensate can interact with the normal atomic cloud through
three-body interaction, which can be phenomenologically incorporated
by a gain/loss term $\eta(t)$.

To reduce Eq.(\ref{GP}) to an effective one dimensional equation, we assume that, the interaction energy of atoms
is much less than the kinetic energy in the transverse direction \cite{Salasch}:
\bea\Psi({\bf{r}},t)=\frac{1}{{\sqrt{2\pi
a_{B}}}a_{\perp}}\psi\left(\frac{z}{a_{\perp}},\omega_{\perp}t\right)
\nonumber
\\\times \exp\left(-i\omega_{\perp}t-\frac{x^2+y^2}{2a_{\perp}^{2}}\right)~.\eea
In dimensionless units, GP equation then reduces to the following
one dimensional nonlinear \SCR equation: \bea \label{NLSE}
i\partial_{t}\psi=-\frac{1}{2}\partial_{zz}\psi+\gamma(t)|\psi|^2
\psi+\frac{1}{2}M(t)z^2\psi+i\frac{g(t)}{2}\psi~.\eea Here,
$\gamma=2a_{s}(t)/a_B$, $M(t)=\omega_{0}^{2}(t)/\omega_{\perp}^{2}$,
$g(t)=\eta(t)/\hbar \omega_{\perp}$,
$a_{\perp}=(\hbar/m\omega_{\perp})^{1/2}$ and $a_B$ is the Bohr's
radius. For the sake of generality we have kept $M(t)$ time
dependent, a constant $M(t)$ implies an oscillator potential which
can be confining or expulsive for $M>0$ or $M<0$, respectively.

In order to discuss the cases of a variety of experimentally
achievable profiles of the time dependent trapping potential and to
obtain corresponding analytical solutions, we assume the following
ansatz solution: \be\label{ansatz}
\psi(z,t)=\sqrt{A(t)}F[A(t)\left\{z-l(t)\right\}]\exp[i\Phi(z,t)+G(t)/2],\ee
where, $G(t)=\int_{0}^{t}g(t')dt'$, $l(t)=\int_{0}^{t}v(t')dt'$ and
we assume that phase has a quadratic form: \be\label{Chirp}
\Phi(z,t)=a(t)+b(t)z-\frac{1}{2}c(t)z^2.\ee In the above equation, $
a(t)=a_{0}+\frac{\lambda-1}{2}\int_{0}^{t}A^{2}(t')dt'$, and $c(t)$
is determined by a Riccati type equation:

\be\label{Rica} c_{t}-c^{2}(t)=M(t). \ee Interestingly, this
equation can be expressed as a \SCR eigenvalue problem via a change
of variable, $c(t)=-\frac{d\ln[\varphi(t)]}{dt}$: \be
-\varphi''(t)-M(t)\varphi(t)=0.\ee Taking advantage of this connection, below we show that,
corresponding to each solvable
quantum mechanical system, one can identify a soliton configuration.
The fact that, \SCR equation can be exactly solved for a variety
of $V(t)$, gives us freedom to control the dynamics of BEC in a
number of analytically tractable ways. This is one of the main
results of this Letter. Although a host of time dependent oscillator
frequencies can be addressed, we will only demonstrate in
the text, a few experimentally realizable examples.

We also find the following consistency conditions

\begin{equation}\label{consi}
\left.
\begin{array}{c}
\gamma(t)=\gamma_{0}e^{-G(t)/2}A(t)/A_{0},~~b(t)=A(t) \\
A(t)=A_{0}\exp\left\{ \int_{0}^{t}c(t')dt'\right\},~~A_{0}>0\\
\frac{dl(t)}{dt}-c(t) l(t)=b(t)
\end{array}
\right\}
\end{equation}

Now substitution of the ansatz \eqrf{ansatz} in \eqrf{NLSE} and
using the consistency conditions Eq.(\ref{consi}), we obtain the
differential equation for the function $F$ in terms of the new
variable $T=A(t)[z-l(t)]$:

\be \label{elliptic} F''(T)-\lambda F(T)+2\kappa F^{3}(T)=0,
~{\mathrm{where}}~ \kappa=-\frac{\gamma_{0}}{A_0}. \ee

We note that, the Eq. (\ref{elliptic}) possesses a variety of
solutions in the form of twelve Jacobian elliptic functions
{\it{e.g.,}} $\mathrm{cn}(T,m)$, $\mathrm{sn}(T,m)$ and
$\mathrm{dn}(T,m)$ etc. Here $m$ is the modulus parameter taking
value in the interval $0\le m\le 1$. These functions interpolate
between the trigonometric and hyperbolic functions in the limiting
cases $m=0$ and $1$, respectively. In the appropriate parameter
regimes, a wide variety of solutions emerge which include soliton
trains, akin to experimental observations of Strecker {\it{et al.,}}
and localized bright and dark solitons.

The bright soliton trains of the form
$\psi(z,t)=\sqrt{A(t)}{\mathrm{cn}}\left[T/\tau_{0},m\right]\exp[i\Phi(z,t)+G(t)/2]$
exist for $\kappa>0$, where
$\tau_{0}^{2}=-A_{0}(m+m^{2})/2\gamma_{0}$ and
$\lambda=(m^{2}+m-1)/2\tau_{0}^{2}$. Here, all the coefficients are
determined from Eqs. (\ref{Chirp}-\ref{consi}). It should be pointed
out that, in the repulsive domain $\gamma_{0}>0$ the localized dark
soliton is of the form
$\psi(z,t)=\sqrt{A(t)}\tanh\left[\frac{A(t)z}{\tau_{0}}\right]\exp[i\Phi(z,t)+G(t)/2],$
since the same vanishes at the origin. The cnoidal wave solution for
this is of $\mathrm{sn}(T,m)$ type.

Below, we examine the formation and dynamics of nonlinear
excitations in the presence of an oscillator potential, with a
variety of experimentally achievable temporal modulations in the
frequency.

\noindent{\it{Soliton trains in a confining oscillator.---}}
Inspired by the experiments of Strecker {\it{et.al,}}
\cite{Strecker} in the presence of confining oscillator, with no
gain/loss we obtain following moving soliton trains with velocity
$v(t)=A_{0}\cos(M_{0}t)$:

\bea
\psi(z,t)&=&\sqrt{A_{0}\sec(M_{0}t)}\nonumber\\&\times&{\mathrm{cn}}\left[\frac{A_{0}\sec(M_{0}t)
(z-\frac{A_{0}}{M_{0}}\sin[M_{0}t])}
{\tau_{0}},m\right]e^{i\Phi}.\eea

\begin{figure}
\includegraphics[width=1.7in]{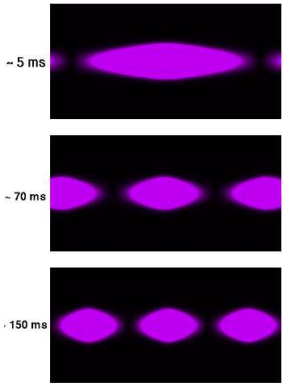}
\caption{Plots depicting the snapshots of soliton trains at times
$\approx 5 ms$, $\approx 70 ms$ and $\approx 150 ms$ with parameters
$M=.25$, $\gamma=-0.5$, $\tau_{0}=.9999$, $A_{0}=0.5$ and modulus
parameter $m=.99999$.}\label{Trains}
\end{figure}

As shown in Fig. (\ref{Trains}) initially at $\approx 5 ms$ there is
only one profile and as time progresses it breaks up in to many
profiles giving rise to a soliton train.

\begin{figure}
\includegraphics[width=2.25in]{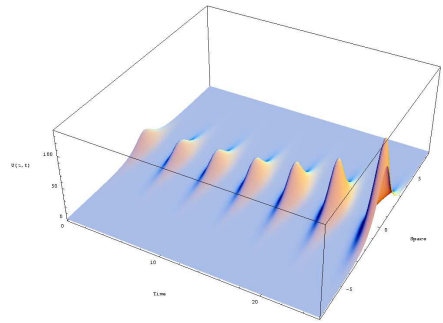}
\caption{Figure depicting collapse and revival of atomic condensate
and amplification through periodic exchange of atoms with the
background, for the parameter values $M=0.0025$, $\tau_{0}=2.5$,
$A_{0}=3.125$, $\gamma_{0}=-0.5$ as in the Ref. \cite{Strecker} and
gain parameters $a_{1}=.01$, $a_{2}=8.25$ and $\delta=1.5$.
\label{BEC2}}
\end{figure}

\noindent{\it{Collapse and revival of the condensate.---}} Inspired
by the experiments of Strecker {\it{et al.,}} we tailor the gain
profile in the same range of the parameters, where one observes
dramatic collapse and revival of condensates with an increase in the
amplitude. Specifically we take, $M(t)=M_{0}^2$,
$g(t)=a_{1}t-a_{2}\sin(\delta t)$ and
$\gamma(t)=\gamma_{0}\exp\left[
\frac{-a_{1}t^2}{2}-\frac{a_{2}}{\delta}\cos(\delta
t)\right]\sec(M_{0}t)$, for which we obtain from \eqrf{consi},
$A(t)=A_{0}\sec(M_0t)$. The complete solution can be written as \be
\psi(z,t)=
\sqrt{A_{0}\sec(M_0t)}\sech\left[\frac{A_{0}\sec(M_0t)z}{\tau_{0}}
\right]\exp\left[G(t)/2\right]e^{i\Phi}. \ee  This is depicted in
Fig.(\ref{BEC2}). In this case, sinusoidal nature of gain function
implies a periodic exchange of atoms with the background. In the
attractive domain, presence of a background surrounding the
condensate has been seen in the experiments leading to formation of
bright solitons \cite{Strecker,Khaykovich}. In this light, the
collapse and revival of the condensate, having a sinusoidal exchange
of atoms with the background, may be amenable for verification.

\noindent{\it{Soliton formation from step changes in the trap
potential.---}} Below, we explicate in the attractive domain
($\gamma_{0}<0$) the effect of a step like change in the oscillator
frequency which is well mimicked by the function
$M(t)=-4+\frac{3}{2}[1+\tanh(t/2)]$, for which the coupling has the
form $\gamma(t)=\gamma_{0}\left[\frac{e^{2t}}{(1+e^t)}\right]$,
$\gamma_{0}<0$.
As time progresses, the coupling strength changes exponentially and
the condensate wave function starts building up rapidly. In this
expulsive case, like the experimentally observed bright soliton
\cite{Khaykovich}, one obtains, \bea
\label{BTanh}\psi(z,t)={\sqrt{\frac{A_0e^{2t}}{(1+e^t)}}}
\sech\left[\frac{A_0e^{2t}z}{(1+e^t)\tau_{0}}\right]\exp[i\Phi(z,t)].
\eea

\begin{figure}
\includegraphics[width=2.25in]{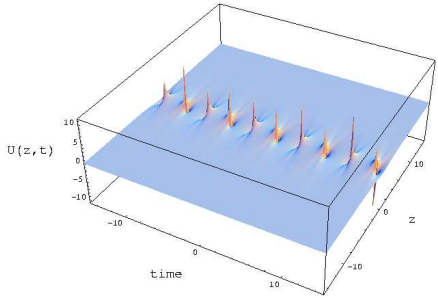}
\caption{Figure depicting spatio-temporal pattern of condensate wave
function in a confining oscillator potential ($M(t)=0.25$) with
constant attractive coupling $\gamma(t)=-0.5$ and $A_{0}=0.5$,
$\tau_{0}=1$.\label{spatio-temp}}
\end{figure}

\noindent{\it{New solutions in the case of constant coupling.---}} A
number of interesting condensate profiles emerge in the constant
attractive coupling regime, depending on the nature of the external
potential. For the regular oscillator confinement one obtains a
spectacular spatio-temporal pattern given by Eq. (\ref{SPTP}) in the
amplitude of the order parameter, as seen in Fig.
(\ref{spatio-temp}). The extreme increase in amplitude happens due
to the presence of the $\sec(M_{0}t)$ in the amplitude.
\be\label{SPTP}
\psi(z,t)=\sqrt{A_{0}\sec(M_{0}t)}\sech\left[\frac{A_{0}\sec(M_{0}t)}{\tau_{0}}
\right]\exp[G/2]e^{i\Phi} \ee

\begin{figure}
\includegraphics[width=2.25in]{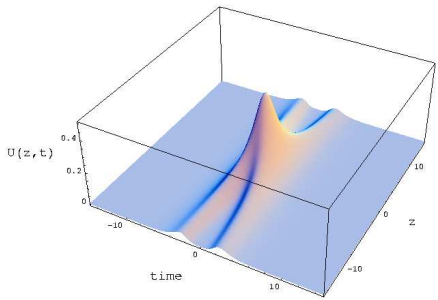}
\caption{Formation of two-soliton bound state in an expulsive
potential($M(t)=-0.25$) with all the other parameters having same
values as in Fig.(\ref{spatio-temp}).} \label{soliton-antisoliton}
\end{figure}

\noindent{\it{New bound state for solitons.---}}Interestingly, as
shown in Fig. (\ref{soliton-antisoliton}), in the expulsive domain
one obtains bound states of solitons in the same parameter regime.
The analytical expression for this remains same as in
Eq.(\ref{SPTP}), except for the value of $A(t)$ which is now given
by$ A(t)=A_{0}\sech[M_{0}t]$. It is worth mentioning that, in
optical fibers with a variable dispersion similar structures have
been seen through numerical investigations \cite{Kumar}.

Keeping in mind, the nontrivial nature of the phase and its
connection with soliton profile, the nature of the spatio-temporal
dynamics of the same for various configurations is worth discussing.
The quadratic nature of the phase with respect to the space
coordinate introduces a spatial chirping effect. Furthermore, its
coefficient $c(t)$ is also time-dependent and is connected with the
external potential through Eq. (\ref{Rica}); hence the trap
potential has a signature on the same. The purely temporal part
$a(t)$ is also sensitive to trap potential as depicted by Eq.
(\ref{consi}).

In conclusion, for one dimensional GP equation in an oscillator
potential, with time dependent coupling and gain/loss, we have
obtained a wide class of exact solutions. These include
experimentally observed bright soliton profile in the attractive
coupling regime for the expulsive case. The fact that the equation
governing the soliton profiles in the presence of time dependent
harmonic oscillator, maps to the \SCR equation, opens up a host of
opportunities. Corresponding to any solvable quantum mechanical
potential, a time dependent cigar-shaped BEC profile can be
obtained. We have explicated amplification of BEC profile through
smooth variation of the oscillator frequency. Formation of soliton
bound states and spectacular spatio-temporal patterns that can
manifest in this nonlinear system with time-dependent control
parameters are demonstrated. One observes dramatic compression and
localization of broad condensate profiles, which may have
technological implications. We have analyzed the nonlinear
excitations in the presence of oscillator potential having linear,
quadratic, Morse and other type of time dependencies.

\end{document}